\documentclass[twocolumn,showpacs,preprintnumbers,superscriptaddress,amsmath,amssymb,pre]{revtex4}

\usepackage{graphicx}
\usepackage{dcolumn}
\usepackage{bm}

\begin{document}

\preprint{APS/Shape complexity}
\title{Shape complexity and fractality of fracture surfaces of swelled isotactic polypropylene with supercritical carbon dioxide}
\author{Wei-Xing Zhou}
\email{wxzhou@moho.ess.ucla.edu}
\author{Bin Li}
\author{Tao Liu}
\author{Gui-Ping Cao}
\author{Ling Zhao}
\author{Wei-Kang Yuan}
\affiliation{State Key Laboratory of Chemical Reaction
Engineering,\\ East China University of Science and Technology,
Shanghai 200237, China}

\date{\today}

\begin{abstract}
We have investigated the fractal characteristics and shape
complexity of the fracture surfaces of swelled isotactic
polypropylene Y1600 in supercritical carbon dioxide fluid through
the consideration of the statistics of the islands in binary SEM
images. The distributions of area $A$, perimeter $L$, and shape
complexity $C$ follow power laws $p(A)\sim A^{-(\mu_A+1)}$,
$p(L)\sim L^{-(\mu_L+1)}$, and $p(C)\sim C^{-(\nu+1)}$, with the
scaling ranges spanning over two decades. The perimeter and shape
complexity scale respectively as $L\sim A^{D/2}$ and $C\sim A^q$
in two scaling regions delimited by $A\approx 10^3$. The fractal
dimension and shape complexity increase when the temperature
decreases. In addition, the relationships among different
power-law scaling exponents $\mu_A$, $\mu_B$, $\nu$, $D$, and $q$
have been derived analytically, assuming that $A$, $L$, and $C$
follow power-law distributions.
\end{abstract}


\maketitle

\section{Introduction}
\label{sec:intro}

Polypropylene, being one of the fastest growing engineering
plastics, has wide industrial and everyday life applications due
to its intrinsic properties such as low density, high melting
point, high tensile strength, rigidity, stress crack resistance,
abrasion resistance, low creep and a surface which is highly
resistance to chemical attack, and so on \cite{Moore-1996}.
Impregnation of nucleating agents allows the polymer to be
crystallized at a higher temperature during processing and changes
a lot the mechanical and optical performance of the resultant
polymer matrix \cite{Xu-Lei-Xie-2003-MD,Iroh-Berry-1996-EPJ}. In
such nucleating agent impregnating processes, supercritical fluids
are well-established swelling solvents whose strength can be tuned
continuously from gas-like to liquid-like by manipulating the
temperature and pressure. Especially, supercritical carbon dioxide
is a good swelling agent for most polymers and will dissolve many
small molecules
\cite{Chiou-Barlow-Paul-1985-JAPS,Kamiya-Hirose-Naito-Mizoguchi-1988-JPSB,Berens-Huvard-Korsmeyer-Kunig-1992-JAPS,Kamiya-Mizoguchi-Terada-Fujiwara-Wang-1998-MM},
except for some fluoropolymers and silicones
\cite{Tuminello-Dee-McHugh-1995-MM}. This provides the ability to
control the degree of swelling in a polymer
\cite{Fleming-Koros-1986-MM,Pope-Koros-1996-JPSB,Condo-Sumpter-Lee-Johnston-1996-IECR}
as well as the partitioning of small molecule penetrants between a
swollen polymer phase and the fluid phase
\cite{Vincent-Kazarian-Eckert-1997-AICHE,Kazarian-Vincent-West-Eckert-1998-JSF}.

The morphology of fracture surfaces of a material is of great
concerns and interests in many studies
\cite{Cherepanov-Balankin-Ivanova-1995-EFM,Fineberg-Marder-1999-PR}.
A frequently used tools is fractal geometry pioneered by
Mandelbrot's celebrated work \cite{Mandelbrot-1983}. Specifically,
fractal geometry has been widely applied to the topographical
description of fracture surfaces of metals
\cite{Mandelbrot-Passoja-Paullay-1984-Nature,Wendt-SL-Smid-2002-JM},
ceramics
\cite{Mecholsky-Passoja-FR-1989-JACS,Thompson-Anusavice-Balasubramaniam-Mecholsky-1995-JACS,Celli-Tucci-Esposito-Palmonari-2003-JECS},
polymers \cite{Chen-Runt-1989-PC,Yu-Xu-Tian-Chen-Luo-2002-MD},
concretes
\cite{Dougan-Addison-2001-CCR,Wang-Diamond-2001-CCR,Issa-Issa-Islam-Chudnovsky-2003-EFM,Yan-Wu-Zhang-Yao-2003-CCC},
alloys
\cite{Shek-Lin-Lee-Lai-1998-JNCS,Wang-Zhou-Wang-Tian-Liu-Kong-1999-MSEA,Betekhtin-Butenko-Gilyarov-Korsukov-Lukyanenko-Obidov-Khartsiev-2002-TPL,Eftekhari-2003-ASS},
rocks
\cite{Xie-Sun-Ju-Feng-2001-IJSS,Babadagli-Develi-2003-TAFM,Zhou-Xie-2003-SRL},
and many others. There are many different approaches adopted in
the determination of fractal dimensions of fracture surfaces and
the estimated fractal dimensions from different materials vary
from case to case and are not universal \cite{Wang-2004-PB}.

Concerning the fractal characterization of the fracture surfaces
of polypropylene, it was found that the fractal dimension of the
fracture surfaces is less than 2.12
\cite{Yu-Xu-Tian-Chen-Luo-2002-MD}. In this paper, we investigate
the shape complexity and fractality of fracture surfaces of
nucleating agents impregnated isotactic polypropylene (Y1600)
swelled with supercritical carbon dioxide based on the
area-perimeter relationship
\cite{Mandelbrot-1983,Lovejoy-1982-Science,Mandelbrot-Passoja-Paullay-1984-Nature}.
We shall see that the fractal dimension of the fracture surface of
supercritical dioxide swelled polypropylene is much higher than
that without swelling and decreases with increasing temperature.

\section{Experimental}

The isotactic polypropylene (Y1600) powder we have used has an
average diameter of $3\sim4~mm$. Fifteen grams of isotactic
polypropylene powder was melt in an oven at
$200\,^{\circ}\mathrm{C}$ and then made into film with a thickness
of 0.3 $mm$ by using a press machine with a pressure of 0.5 GN.
The size of the standard film samples was $1mm\times 3mm$ and are
refluxed with acetone for twenty four hours to remove the
impurities, and then annealed at $200\,^{\circ}\mathrm{C}$ for two
hours. The nucleating agent we used was NA21.

\begin{figure}
\begin{center}
\includegraphics[width=8cm, height=6cm]{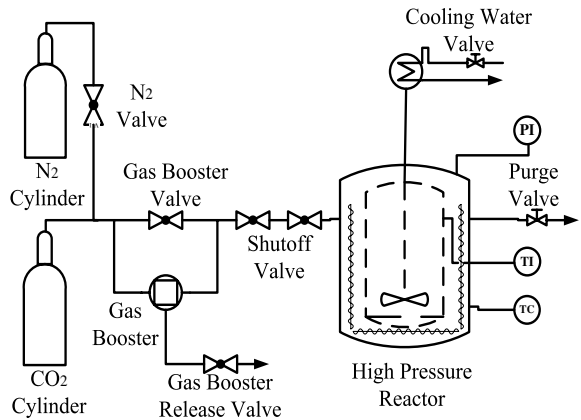}
\end{center}
\caption{Schematic flow chart of the experimental apparatus.}
\label{FigSchDiag}
\end{figure}

The schematic flow chart of the experimental apparatus is
illustrated in Fig.~\ref{FigSchDiag}. The experimental apparatus
consists mainly of a gas cylinder, a gas booster, a digital
pressure gauge, an electrical heating bath, and valves and
fittings of different kinds. The system was cleaned thoroughly
using suitable solvents and dried under vacuum. Isotactic PP films
marked from one to twelve were placed in the high pressure cell
together with desired amount ($1\%$ wt) of nucleating agent
(NA21). The system was purged with $\rm{CO}_2$ and after the
system had reached the desired temperature and thermal
equilibrium, $\rm{CO}_2$ was charged until the desired pressure
was reached. The impregnation process lasted for four hours and
then depressurized the $\rm{CO}_2$ from the high pressure cell
rapidly. Then the vessel was cooled and opened, and the specimens
were taken out for analysis. All of the metallic parts in contact
with the studied chemicals were made of stainless steel. The
apparatus was tested up to 35 MPa. The total volume of the system
is 500 mL.

\begin{figure}
\begin{center}
\includegraphics[width=8cm, height=6cm]{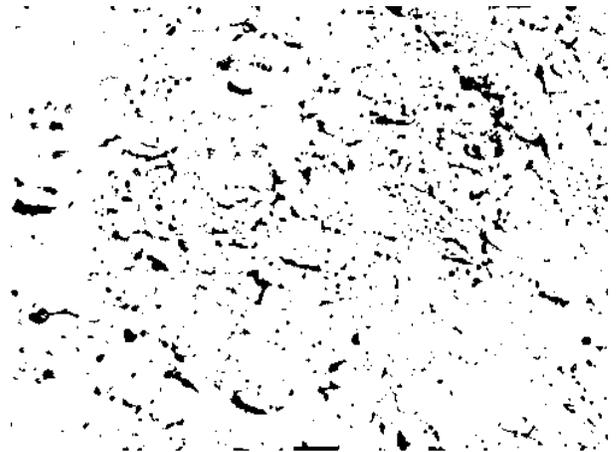}
\end{center}
\caption{Image of the recognized islands from a typical SEM
picture of swelled isotactic polypropylene impregnated with NA21
at level set of 0.7.} \label{FigIslands}
\end{figure}

The experiments were performed at fixed pressure of 7.584 MPa and
at four different temperatures of $T=60\,^{\circ}\mathrm{C}$,
$80\,^{\circ}\mathrm{C}$, $100\,^{\circ}\mathrm{C}$, and
$120\,^{\circ}\mathrm{C}$. For each sample, we took ten SEM
pictures at magnification of 5000 or 10000. The pictures have 256
grey levels. Due to the nature of area-perimeter approach, the
results are irrelevant to the magnification
\cite{Mandelbrot-1983}. A grey-level picture was transformed into
black-and-white image for a given level set $\epsilon$ according
to the criterion that a pixel is black if its grey level is larger
than $256\epsilon$ and is white otherwise. The resultant binary
image has many black islands. A typical binary image of islands is
shown in Fig.~\ref{FigIslands}. In our calculations, we have used
eleven level sets from 0.45 to 0.95 spaced by 0.05 for each SEM
picture.

\section{Fractality and shape complexity}
\label{s1:ShapeComplexity}

\subsection{Rank-ordering statistics of area and perimeter}
\label{s2:RankOrdering}

In order to estimate the probability distribution of a physical
variable empirically, several approaches are available. For a
possible power-law distribution with fat tails, cumulative
distribution or log-binning technique are usually adopted. A
similar concept to the complementary distribution, called
rank-ordering statistics \cite{Sornette-2000}, has the advantage
of easy implementation, without loss of information, and less
noisy.

Consider $n$ observations of variable $x$ sampled from a
distribution whose probability density is $p(x)$. Then the
complementary distribution is $P(y > x) = \int_x^\infty p(y) dy$.
We sort the $n$ observations in non-increasing order such that
$x_1\ge x_2 \ge \cdots \ge x_R \ge \cdots \ge x_n$, where $R$ is
the rank of the observation. It follows that $nP(x \ge x_R)$ is
the expected number of observations larger than or equal to $x_R$,
that is,
\begin{equation}
 nP(x \ge x_R) = R~.
 \label{Eq:nPR}
\end{equation}
If the probability density of variable $x$ follows a power law
that $p(x) \sim x^{-(1+\mu)}$, then the complementary distribution
$P(x) \sim x^{-\mu}$. An intuitive relation between $x_R$ and $R$
follows
\begin{equation}
 x_R \sim R^{-1/\mu}~.
 \label{Eq:xR}
\end{equation}
A rigorous expression of (\ref{Eq:xR}) by calculating the most
probable value of $x_R$ from the probability that the $R$-th value
equals to $x_R$ gives \cite{Sornette-2000}
\begin{equation}
 x_R \sim \left(\frac{\mu n+1}{\mu R+1}\right)^{1/\mu}~.
 \label{Eq:xR0}
\end{equation}
when $\mu R\gg 1$ or equivalently $1\ll R\le N$, we retrieve
(\ref{Eq:xR}). A plot of $\ln x_R$ as a function of $\ln R$ gives
a straight line with slope $-1/\mu$ with deviations for the first
a few ranks if $x$ is distributed according to a power law of
exponent $\mu$.

We note that the rank-ordering statistics is nothing but a simple
generalization of Zipf's law
\cite{Zipf-1949,Mandelbrot-1983,Sornette-2000} and has wide
applications, such as in linguistics \cite{Mandelbrot-1954-Word},
the distribution of large earthquakes
\cite{Sornette-Knopoff-Kagan-Vanneste-1996-JGR}, time-occurrences
of extreme floods \cite{Mazzarella-Rapetti-2004-JH}, to list a
few. More generally, rank-ordering statistics can be applied to
probability distributions other than power laws, such as
exponential or stretched exponential distributions
\cite{Laherrere-Sornette-1998-EPJB}, normal or log-normal
distributions \cite{Sornette-2000}, and so on.

\begin{figure}
\begin{center}
\includegraphics[width=8cm, height=6cm]{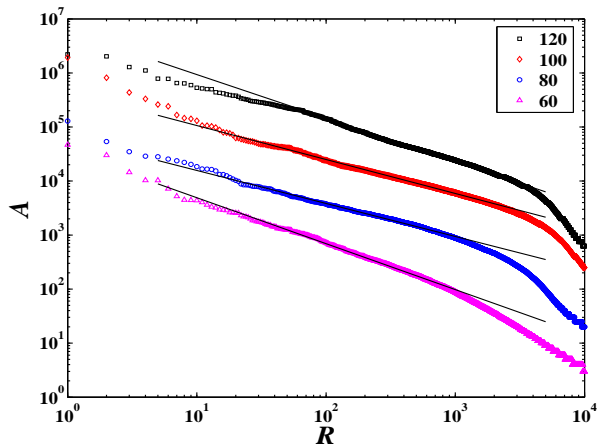}
\end{center}
\caption{Log-log plot of the rank-ordered areas $A$ at different
temperatures shown in the legend. The plots are translated
vertically for clarity. The solid lines are linear fits to the
data at $59\le R\le 3890$ for $T=120\,^{\circ}\mathrm{C}$, $10\le
R\le 1995$ for $T=100\,^{\circ}\mathrm{C}$, $22\le R\le 1000$ for
$T=80\,^{\circ}\mathrm{C}$, and $8\le R \le 800$ for
$T=60\,^{\circ}\mathrm{C}$, respectively.} \label{FigRA}
\end{figure}

Figure \ref{FigRA} illustrates the log-log plots of the
rank-ordered areas $A$ at different temperatures. There is clear
power-law dependence between $A$ and its rank $R$ with the scaling
ranges spanning over about two decades. Least-squared linear
fitting gives $1/\mu_A=0.80\pm 0.05$, $1/\mu_A=0.63 \pm 0.02$,
$1/\mu_A=0.61 \pm 0.03$, and $1/\mu_A=0.85 \pm 0.04$ with
decreasing temperatures, where the errors are estimated by the
r.m.s. of the fit residuals. Using $\sigma_{\mu_A} =
\sigma_{1/\mu_A}/\mu_A^2$, we obtain the power-law exponents
$\mu_A= 1.24\pm 0.08$, $\mu_A=1.59\pm 0.06$, $\mu_A=1.64\pm 0.09$,
and $\mu_A=1.18\pm 0.05$ with decreasing temperatures.

\begin{figure}
\begin{center}
\includegraphics[width=8cm, height=6cm]{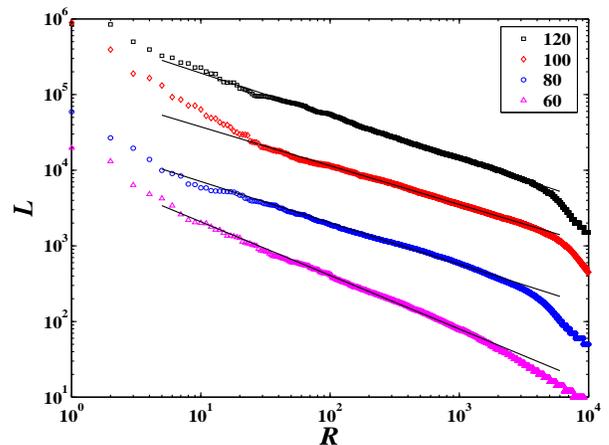}
\end{center}
\caption{Log-log plot of the rank-ordered perimeters $L$ at
different temperatures marked in the legend. The plots are
translated vertically for clarity. The solid lines are linear fits
to the data at $5\le R\le 3890$ for $T=120\,^{\circ}\mathrm{C}$,
$24\le R\le 3000$ for $T=100\,^{\circ}\mathrm{C}$, $16\le R\le
2000$ for $T=80\,^{\circ}\mathrm{C}$, and $10\le R \le 1500$ for
$T=60\,^{\circ}\mathrm{C}$, respectively.} \label{FigRL}
\end{figure}

Figure \ref{FigRL} shows the log-log plots of the rank-ordered
perimeters $L$ at different temperatures. There are also clear
power-law relations between $L$ and its rank $R$ whose scaling
ranges spanning over about three decades. Least-squared linear
fitting gives $1/\mu_L=0.56\pm 0.02$, $1/\mu_L=0.51 \pm 0.01$,
$1/\mu_L=0.55 \pm 0.03$, and $1/\mu_L=0.71 \pm 0.03$ with
decreasing temperatures. We thus obtain the power-law exponents
$\mu_L= 1.78\pm 0.07$, $\mu_L=1.95\pm 0.05$, $\mu_L=1.83\pm 0.12$,
and $\mu_L=1.41\pm 0.06$ with decreasing temperatures. It is
noteworthy that the scaling ranges of the rank-ordering of both
$A$ and $P$ are well above two decades broader than most of other
experiments with scaling ranges centered around 1.3 orders of
magnitude and spanning mainly between 0.5 and 2.0
\cite{Malcai-Lidar-Biham-Avnir-1997-PRE,Avnir-Biham-Lidar-Malcai-1998-Science}.

\subsection{Scaling between area and perimeter} \label{s2:AL}

If an island is fractal, its area $A$ and perimeter $L$ follow a
simple relation,
\begin{equation}
 L^{1/D} \sim A^{1/2}~,
 \label{Eq:LA}
\end{equation}
where $D$ is the fractal dimension describing the wiggliness of
the of the perimeter. For $D=1$, the perimeter of the island is
smooth. For $D=2$, the perimeter becomes more and more contorted
to fill the plane. This area-perimeter relation (\ref{Eq:LA}) can
also be applied to many islands when they are self-similar in the
sense that the ratio of $L^{1/D}$ over $A^{1/2}$ is constant for
different islands \cite{Mandelbrot-1983}. It has been used to
investigate the geometry of satellite- and radar-determined cloud
and rain regions by drawing $A$ against $L$ for different regions
in a log-log plot \cite{Lovejoy-1982-Science}. The slope of the
straight line obtained by performing a linear regression of the
data gives $2/D$.

Following the same procedure, we plotted for each temperature $L$
against $A$  for all islands identified at different level sets.
The resultant scatter plot for a given temperature shows nice
collapse to a straight line. The four slopes are $D=1.34 \pm 0.08$
for $T=60\,^{\circ}\mathrm{C}$, $D=1.32 \pm 0.07$ for
$T=80\,^{\circ}\mathrm{C}$, $D=1.31 \pm 0.08$ for
$T=100\,^{\circ}\mathrm{C}$, and $D=1.28 \pm 0.09$ for
$T=120\,^{\circ}\mathrm{C}$, respectively. The errors were
estimated by the r.m.s. of the fit residuals.

\begin{figure}
\begin{center}
\includegraphics[width=8cm, height=6cm]{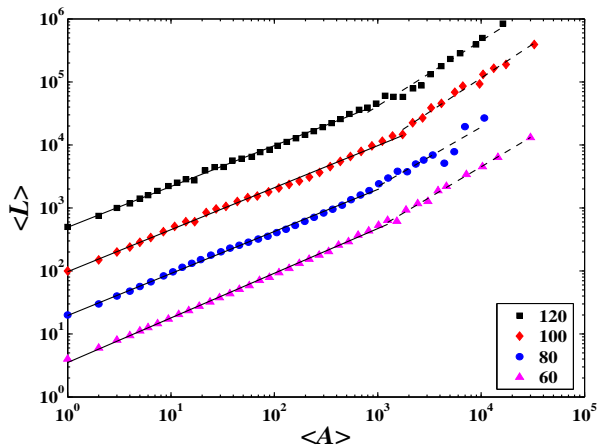}
\end{center}
\caption{Plots of $\langle{L}\rangle$ with respect to
$\langle{A}\rangle$ at different temperatures. The plots are
translated vertically for clarity. We note that
$\langle{L}\rangle=4$ when $\langle{A}\rangle=1$ in all the four
plots.} \label{FigLA}
\end{figure}

However, a closer investigation of the scatter plots shows that
there are two scaling regions with a kink at around $A=1000$. To
have a better view angle, we adopt an averaging technique for both
$A$ and $L$. We insert $n-2$ points in the interval $[\min(A),
\max(A)]$ resulting in $\min(A)=a_0<a_1<a_2<\cdots<a_n=\max(A)$ so
that $a_i$'s are logarithmically spaced. We can identify all
islands whose areas fall in the interval $[a_i,a_{i+1})$. The
geometric means of $L$ and $A$ are calculated for these islands,
denoted as $\langle{L}\rangle$ and $\langle{A}\rangle$. The
calculated means $\langle{L}\rangle$ and $\langle{A}\rangle$ are
plotted in Fig.~\ref{FigLA}. The two scaling regions are fitted
respectively with two straight lines and we have $D_1=1.42 \pm
0.06$ for $T=60\,^{\circ}\mathrm{C}$, $D_1=1.33 \pm 0.06$ for
$T=80\,^{\circ}\mathrm{C}$, $D_1=1.33\pm 0.07$ for
$T=100\,^{\circ}\mathrm{C}$, and $D_1=1.30\pm 0.05$ for
$T=120\,^{\circ}\mathrm{C}$,  in the first region that
$\langle{A}\rangle < 1000$ and $D_2=1.96 \pm 0.07$ for
$T=60\,^{\circ}\mathrm{C}$, $D_2=1.89\pm 0.24$ for
$T=80\,^{\circ}\mathrm{C}$, $D_2=2.15\pm 0.11$ for
$T=100\,^{\circ}\mathrm{C}$, and $D_2=2.07 \pm 0.14$ for
$T=120\,^{\circ}\mathrm{C}$ in the second region that
$\langle{A}\rangle>1000$.

It is very interesting to notice that $L \propto A$ for large
areas. Under the consumption that the islands are fractals, this
relation can be interpreted that the perimeters of the large
islands are so wiggly that they can fill the plane. However, there
are simple models that can deny this consumption of
self-similarity. Consider that we have a set of strip-like islands
with length $l$ and width $w$. We have $L = 2(l+w)$ and $A=lw$.
When $w$ is fixed and $l \gg w$, it follows that $L \propto A$.
This simple model can indeed explain the current situation. For
cracked polypropylene, there are big strip-like ridges in the
intersection whose surface is relatively smooth. For small
islands, the fractal nature is more sound.

According to the additive rule of codimensions of two intersecting
independent sets, the fractal dimension of the surface is $D_s =
D_1 +1$ \cite{Mandelbrot-1983}. Therefore, $D_s=2.42 \pm 0.06$ for
$T=60\,^{\circ}\mathrm{C}$, $D_s=2.33 \pm 0.06$ for
$T=80\,^{\circ}\mathrm{C}$, $D_s=2.33\pm 0.07$ for
$T=100\,^{\circ}\mathrm{C}$, and $D_s=2.30\pm 0.05$ for
$T=120\,^{\circ}\mathrm{C}$. We see that the fractal dimension
$D_s$ increases with decreasing temperature. In other words, with
the increase of temperature, the surface becomes smoother.

Assuming that $p(A)\sim A^{-(\mu_A+1)}$ and $p(L)\sim
L^{-(\mu_L+1)}$, we can estimate the fractal dimension $D$ in
(\ref{Eq:LA}) according to the relation $p(A)dA = p(L)dL$ such
that
\begin{equation}
 D=2\mu_A/\mu_L~.
 \label{Eq:Dmu}
\end{equation}
The four estimated values of $D$ using this relation
(\ref{Eq:Dmu}) are 1.67, 1.79, 1.63, and 1.40 with increasing
temperature. The discrepancy from $D_1$ is remarkable ($15\%$,
$25\%$, $18\%$, and $7\%$, accordingly). The source of this
discrepancy is threefold. Firstly, the power-law distributions of
$A$ and $L$ have cutoffs at both ends of small and large values so
that the derivation of (\ref{Eq:Dmu}) is not rigorous. Secondly,
the determination of the scaling ranges may cause errors. Thirdly,
the scaling ranges of the three power laws are not consistent with
each other.

\subsection{Shape complexity}
\label{s2:ShapeComplexity}

To quantify the shape complexity of an irregular fractal object,
besides the fractal dimension, there are other relevant measures
related to the fractal nature
\cite{Catrakis-Dimotakis-1998-PRL,Kondev-Henley-Salinas-2000-PRE}.
For a $d$-dimensional hypersphere, its surface and volume are
related by
\begin{equation}
S_{d,sph} = \frac{1}{k_d} V_{d,sph}^{(d-1)/d}~, \label{Eq:SV}
\end{equation}
where
\begin{equation}
k_d = \frac{\Gamma^{1/2}(1+d/2)}{d\pi^{1/2}}~. \label{Eq:kd}
\end{equation}
Since a hypersphere has the maximal enclosed volume among the
objects with a given surface $S_d$, we have
\begin{equation}
S_d \ge \frac{1}{k_d} V_{d}^{(d-1)/d}~.
\end{equation}
Then the following dimensionless ratio
\begin{equation}
C_d \triangleq \frac{k_dS_d}{V_{d}^{(d-1)/d}}
\end{equation}
describes the irregularity or complexity of the investigated
object \cite{Catrakis-Dimotakis-1998-PRL}. It follows that $1 \le
C_d \le \infty$. The limits are reached at $C_d = 1$ for
hyperspheres and $C_d = \infty$ for fractals. For real fractal
objects, sometimes known as prefractals
\cite{Addison-2000-Fractals}, the scaling range is not infinite.
The shape complexity $C_d$ is thus finite. In the case of $d=2$,
we have
\begin{equation}
1\le C \triangleq \frac{L}{2\pi^{1/2}A^{1/2}}\le \infty~.
\label{Eq:C}
\end{equation}
This dimensionless measure of shape complexity has been applied to
scalar field of concentration in turbulent jet at high Reynolds
numbers
\cite{Catrakis-Dimotakis-1998-PRL,Catrakis-Aguirre-RP-Thayne-2002-PF}.

\begin{figure}
\begin{center}
\includegraphics[width=8cm, height=6cm]{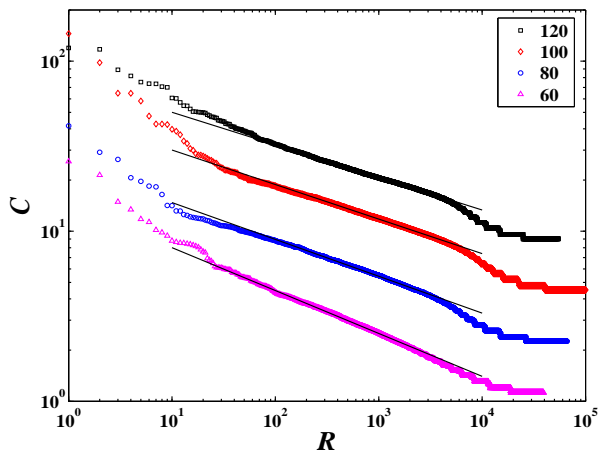}
\end{center}
\caption{Log-log plot of the rank-ordered shape complexity $C$ at
different temperatures. The plots are translated vertically for
clarity. The solid lines are linear fits to the data at $40\le
R\le 3890$ for $T=120\,^{\circ}\mathrm{C}$, $24\le R\le 5000$ for
$T=100\,^{\circ}\mathrm{C}$, $40\le R\le 3000$ for
$T=80\,^{\circ}\mathrm{C}$, and $25\le R \le 1500$ for
$T=60\,^{\circ}\mathrm{C}$, respectively.} \label{FigRC}
\end{figure}

In Fig.~\ref{FigRC} is shown the log-log plot of the rank-ordered
shape complexity $C$ at different temperatures. The power-law
dependence $C \sim R^{-1/\nu}$ implies a power-law probability
density of shape complexity, that is,
\begin{equation}
 p(C) \sim C^{-(\nu+1)}~.
 \label{Eq:pC}
\end{equation}
We find that $\nu=3.97\pm 0.11$ for $T=60\,^{\circ}\mathrm{C}$,
$\nu=4.61\pm 0.17$ for $T=80\,^{\circ}\mathrm{C}$, $\nu=4.93\pm
0.20$ for $T=100\,^{\circ}\mathrm{C}$, and $\nu=5.22\pm 0.17$ for
$T=120\,^{\circ}\mathrm{C}$. The mean logarithmic complexity
$\langle \ln C \rangle = 1/(\nu-1)$
\cite{Catrakis-Dimotakis-1998-PRL} is calculated to be 0.34, 0.28,
0.25, and 0.24 with increasing temperature. In other words, the
shape complexity of the islands decreases with increasing
temperature. This is consistent with the fact the $D$ decreases
with increasing temperature.

\begin{figure}
\begin{center}
\includegraphics[width=8cm, height=6cm]{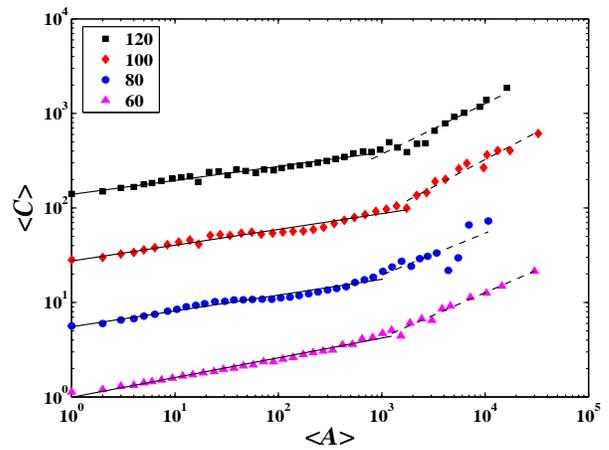}
\end{center}
\caption{Plots of $\langle{C}\rangle$ with respect to
$\langle{A}\rangle$ at different temperatures. The plots are
translated vertically for clarity. We note that
$\langle{C}\rangle=2/\pi^{1/2}$ when $\langle{A}\rangle=1$ in all
the four plots.} \label{FigCA}
\end{figure}

Following the procedure in Sec.~\ref{s2:AL}, we calculate the
means, $\langle{A}\rangle$ and $\langle{C}\rangle$. The results
are shown in Fig.~\ref{FigCA}. Again, we see two scaling regions
guided by straight lines fitted to the data
\begin{equation}
 C \sim A^q~. \label{Eq:CA}
\end{equation}
The slopes are respectively $q_1=0.21$ for
$T=60\,^{\circ}\mathrm{C}$, $q_1=0.17$ for
$T=80\,^{\circ}\mathrm{C}$, $q_1=0.17$ for
$T=100\,^{\circ}\mathrm{C}$, and $q_1=0.15$ for
$T=120\,^{\circ}\mathrm{C}$,  in the first region that
$\langle{A}\rangle < 1000$ and $q_2=0.54$ for
$T=60\,^{\circ}\mathrm{C}$, $q_2=0.58$ for
$T=80\,^{\circ}\mathrm{C}$, $q_2=0.44$ for
$T=100\,^{\circ}\mathrm{C}$, and $q_2=0.48$ for
$T=120\,^{\circ}\mathrm{C}$ in the second region that
$\langle{A}\rangle>1000$. We have verified that $q_i=(D_i-1)/2$
with $i=1,2$ holds exactly for all cases, which is nothing but a
direct consequence of the combination of (\ref{Eq:LA}) and
(\ref{Eq:C}).

Similarly, we can derive the relationship among $q$, $D$, $\mu_A$,
and $\nu$ as follows
\begin{equation}
 \nu = 2\mu_A/(D-1) = \mu_A/q~.
 \label{Eq:numuD}
\end{equation}
Again, the discrepancy between the fitted $\nu$ values and those
estimated indirectly from (\ref{Eq:numuD}) is remarkable ($38\%$,
$48\%$, $53\%$, $30\%$), which can also be resorted to similar
reasoning.

\section{Conclusion}

We have investigated the fractal characteristics of the fracture
surfaces of swelled isotactic polypropylene Y1600 impregnated with
nucleating agent NA21 through consideration of the statistics of
the islands in binary images. At a given temperature, the
distributions of area and perimeter of the islands are found to
follow power laws spanning over two decades of magnitude, via
rank-ordering statistics. The well-established power-law scaling
between area and perimeter shows the overall self-similarity among
islands when $A<10^3$. This shows that the fracture surface is
self-similar. For large islands, the fractal dimension is
estimated to be close to 2 which can be explained by the fact that
most of the large islands are strip-shaped. The fracture surface
is rougher at low temperature with larger fractal dimension.

We have also investigated the shape complexity of the fracture
surfaces using a dimensionless measure $C$. The distribution of
shape complexity is also found to follow a power law spanning over
two decades of magnitude. The shape complexity increases when the
island is larger. There are two power-law scaling ranges between
$C$ and $A$ delimited around $A=10^3$, corresponding to the
two-regime area-perimeter relationship. The exponent $q_1$ serves
as an inverse measure of overall shape complexity, which is
observed to increases with temperature. This is consistent with
the change of fractal dimension at different temperatures.

Furthermore, the relationships among different power-law scaling
exponents of the probability distributions ($\mu_A$, $\mu_B$, and
$\nu$), of area-perimeter relation ($D$), and of complexity-area
relation ($q$) have been derived analytically. However, these
relations hold only when the probability distributions of $A$,
$L$, and $C$ follow exactly power laws. In the present case, we
observed remarkable discrepancy between numerical and analytical
results.

\begin{acknowledgements}
The isotactic polypropylene (Y1600) powder was kindly provided by
the Plastics Department of Shanghai Petrochemical Company and the
nucleating agent, NA21, was kindly supplied by Asahi Denka Co,
Ltd. Apparatus. This work was jointly supported by NSFC/PetroChina
through a major project on multiscale methodology (No. 20490200).
\end{acknowledgements}

\bibliography{IPPShapeComplexity}

\end{document}